\documentstyle[11pt]{article}

\begin{document}
\topmargin 0pt
\oddsidemargin 5mm
\setcounter{page}{1}
\begin{titlepage}
\hfill Preprint YERPHI-1539(13)-99

\hfill  5 August
\vspace{2cm}
\begin{center}

{\bf Lepton-Neutron Bound States }\\
\vspace{5mm}
{\large R.A. Alanakyan}\\
\vspace{5mm}
{\em Theoretical Physics Department,
Yerevan Physics Institute,
Alikhanian Brothers St.2,

 Yerevan 375036, Armenia\\}
 {E-mail: alanak@lx2.yerphi.am\\}
\end{center}

\vspace{5mm}
\centerline{{\bf{Abstract}}}

We consider lepton-neutron (and lepton-antineutron ) bound states and resonances which appear
 due to spin-spin ,spin-orbital interactions,neutron polarization by muon.
Our analis is also true for any system which include one
 charge and one neutral particle with finite size e.g.
 $\pi^0\mu^{\pm}$-bound states.We consider also cylindrically symmetric bound states
and resonances  of particles with anomalous magnetic moments.

\vspace{5mm}
\vfill
\centerline{{\bf{Yerevan Physics Institute}}}
\centerline{{\bf{Yerevan 1999}}}

\end{titlepage}

                 {\bf 1.Introduction}

In this article  we consider lepton-neutron (and lepton-antineutron ) bound states
 and resonances which appear
 due to spin-spin and spin-orbital forces between neutron
and lepton (antilepton) neutron polarization by lepton(
because neutron is composite particle with finite size which consist of  particles
with opposite charge and in the field of lepton neutron polarization appear).

At large distances also exist the potentials  from  $\rho^0-\gamma$
(see Appendix B below)
 mixing.Our analis is also true for any system which include one
 charge and one neutral particle with finite size e.g.
 $\pi^0\mu^{\pm}$,$K^0\mu^{\pm}$-bound states.
As we see below however the attraction  of neutron polarization (see Appendix A)  is
much smaller than spin-spin and spin-orbital interaction and
essential only for $\pi$-meson-lepton bound states).

In non-relativistic approximation spin-spin and spin-orbital
interactions, (which depend in general on angles) has the form \cite{LL4}:
\begin{equation}
\label{A5}
V(r)=6\mu_{\mu}\mu_{n}\frac{1}{r^3}(\frac{\vec{\hat{S}}\vec{r}\vec{\hat{S}}\vec{r}}{r^2}-
\frac{1}{3}(\hat{S})^2)+4\pi\mu_{\mu}\mu_{n}
(\frac{7}{3}(\hat{S})^2-2\delta(\vec{r}))
\end{equation}
 however at $J=0$
$\vec{L}=-\vec{S}$) and potential of spin-spin and spin-orbital interactions become
spherically symmetric
(because $\vec{S}\vec{n}=-\vec{L}\vec{n}=0$) interaction:
\begin{equation}
\label{A5}
V_{SS}(r)=6\mu_{\mu}\mu_{n}\frac{1}{r^3}(-\frac{1}{3}(\hat{S})^2)+4\pi\mu_{\mu}\mu_{n}
(\frac{7}{3}(\hat{S})^2-2\delta(\vec{r}))
\end{equation}
\begin{equation}
\label{A5}
V_{LS}(r)=-6\mu_{\mu}\mu_{n}\frac{1}{r^3}L(L+1)
\end{equation}
We have the following possible
quantum numbers of bound states with  $J=0$:$L=S=0$,$L=S=1$.

The potentials (3)-(5) are non-relativistic.For relativistic consideration it is convenient to use
the Dirac equation in the form:
\begin{equation}
\label{A5}
((E-eA_0)^2-(\vec{p}-e\vec{A})^2+e\vec{\Sigma}\vec{H}-ie\vec{\alpha}\vec{E})\psi=0
\end{equation}
where
$\vec{A}=[\vec{r}\sigma_n]F(r),\vec{E}=-\vec{n}A'_0(r),
H_i=\vec{\sigma}_n(-rF'-2F)+rF'(r)\vec{n}(\vec{\sigma_n}\vec{n})$.

In (4) we can put $\vec{p}\vec{A}=0,\vec{A}\vec{p}=0$ because
$\vec{A}=[\vec{r}\sigma_n]F(r)$.We would like to notice that term
$e\vec{A})^2$ is repulsive.

We find the solution in the following form:
\begin{equation}
\label{A5}
\psi^T=(a(r),i(\vec{\sigma} \vec{n})b(r))
\end{equation}

Taking into account (9)(10) we obtain the following system of radial equations:
\begin{equation}
\label{A5}
\hat{T}a(r)-eA_0'b(r)=0
\end{equation}

\begin{equation}
\label{A5}
\hat{T}b(r)+eA_0'a(r)=0
\end{equation}
where
\begin{equation}
\label{A5}
\hat{T}=(E-eA_0)^2-(\vec{p})^2-(e\vec{A})^2+e \frac{1}{2}(S(S+1)-6)(-rF'-2F)-erF'),
\end{equation}
$(\vec{p})^2=-\frac{1}{r^2}\frac{d}{dr}r^2\frac{d}{dr}+\frac{L(L+1)}{r^2}$.
Of course only $S=0,1$ is possible.

From this system of equations we can obtain:
\begin{equation}
\label{A5}
\hat{T}\frac{1}{eA_0'}\hat{T}a(r)+eA_0'a(r)=0
\end{equation}

It was consideration for particle without anomalous magnetic moment (it is actual
e.g. for positronium).

In case of $\pi^0$-lepton(antileton) bound states if for symplicity we suggest that only
dielectric polarization exist (i.e. only $A_0(r)$ is taken into account) we have the same
 equations as in \cite{LL4}, and for radial functions
($\psi=(f(r)\Omega_{jlm},i^{l-l'},\vec{\sigma}\vec{n}\Omega_{jlm}g(r))$) we obtain the
following  equations where however $V(r)$ is potential from polarization of
$\pi^0$-meson (see Appendix A below):
\begin{equation}
\label{A22}
(f'(r)+\frac{1+\kappa}{r}f(r))-(E+m-V(r))g(r)=0
\end{equation}
\begin{equation}
\label{A22}
(g'(r)+\frac{1-\kappa}{r}g(r))+(E-m-V(r))f(r)=0
\end{equation}
where $\kappa=\mp(j+\frac{1}{2})$ if $j=l\pm\frac{1}{2}$.

It is of interest to consider also bound states of composite neutral fermion
(scalar)-charged  particle,e.g. neutron-ion.If $Z\alpha \sim 1$ polarization effects
may be significant.
The Dirac equation for particle with anomalous magnetic moment has the following form:
\begin{equation}
\label{A18}
(\hat{k}-m_n+\mu_n (\vec{\Sigma}\vec{H}-i\vec{\alpha}\vec{E}) )\psi(k)=0,
\end{equation}
where  $\mu_n $-is anomalous magnetic moment of neutron,$\vec{\Sigma}\vec{H}=F_1(r)$ at $J=0$
and $\vec{\alpha}\vec{E}=\vec{\sigma}\vec{n}G(r)$.

For radial equations we obtain the following results:
\begin{equation}
\label{A22}
(f'(r)+(\frac{1+\kappa}{r}-\mu_nG(r))f(r)-(E+m-\mu_nF_1(r))g(r)=0
\end{equation}
\begin{equation}
\label{A22}
(g'(r)+\frac{1-\kappa}{r}+\mu_nG(r))g(r)+(E-m-\mu_nF_1(r))f(r)=0
\end{equation}
where $\kappa=\mp\frac{1}{2})$ (because $J=0$).If mafnetic field is absent ($F_1(r)=0$)
$J$ may be arbitrary.

Thus we have singular potential $\sim r^{-n}$ at $r>>r_n$.At $r<r_n<<\frac{1}{m_e}$
singular behaviour is absent.It mean the presence
of deep levels of $e^{\pm}n$ bound states. In our next paper
we will calculate it numerically.

In case of neutron the effect of neutron polarization is of order $\sim\frac{\alpha^2}{R}$
is essentially smaller than spin-spin and spin-orbital
interaction $\sim\frac{\alpha}{m_eMR^3}$.However in composite scalar-charged particle
bound states (see below) spin-spin and spin-orbital
interaction  are absent and  attraction by polarization is important.

We would like also to stress that the potential is the sum of the
potential from polarization+formfactors effect (formula ()
above)+ of spin-spin interaction and from photon-$\rho$-meson mixing .

It must be noted also exist long-range potential induced by $\pi^0\gamma\gamma$
vertex:

\begin{equation}
\label{A5}
V(q)=e^2g_{\pi n n}\frac
{1}{q^2+m^2_{\pi}}\vec{E}\vec{H}
\end{equation}
where $\vec{E},\vec{H}$ are fields of the lepton.

{\bf Production and decays of $e^{\pm}n$ atoms }

This bound states may be produced after stop of neutrons (antineutrons) in matter.
Resonances may be produced in $e^{\pm}n$-collisions.
One of the decay mode of this bound states and resonances is weak decays:
\begin{equation}
\label{A5}
e^{\pm}n\quad atoms\rightarrow\nu p
\end{equation}
$e^{\pm}\pi^0(K^0)$ atoms may be produced by the same way.

Also exist potential connected with neutron formfactor:
\begin{equation}
\label{A1}
M=eF_n(\vec{q}^2)\bar{n}\gamma_a n e\gamma_b e
\end{equation}
where $F_n(0)=0$.The formfactor correspond to the some charge density and we obtain:
\begin{equation}
\label{A5}
V(r)=-\alpha\int \frac{1}{|\vec{r}-\vec{r}'|}\rho(\vec{r}')d^3r'
\end{equation}
Thus we have singular potential $\sim r^{-n}$ at $r>>r_n$.At $r<r_n<<\frac{1}{m_e}$
singular behaviour is absent. It mean the presence
of deep levels of $e^{\pm}n$ bound states.In our next paper
we will calculate it numerically.

We would like also to stress that the potential is the sum of the
potential from polarization+formfactors effect (formula ()
above)+ of spin-spin interaction and  also vector
boson-photon mixing which fall as $r^{-3}$ and the most essential at large r.

In case of composite scalar particle-charged scalar particle bound states
and resonances the situation is symplified because
$\pi^0$ is spinless.For $e^{\pm}$ may be used Klein-Gordon equation:
\begin{equation}
\label{A5}
((E-V(r))^2+\frac{1}{r^2}\frac{d}{dr}r^2\frac{d}{dr}-\frac{l(l+1)}{r^2}-m^2)\phi(r)=0
\end{equation}
Numerical solution of this equation for has been performed in \cite{I}.
In the same reference has been considered also task on energy levels of electron in the
field of conducting spheres (see below).The dependence of binding energy versus
rudius of the sphere is shown  on the Fig.3 of this reference.

It must be noted that in \cite{I}  has been considered only case
of homogeneous dielectric sphere, in the future we plane to consider more realistic
potential ().

It is of interest to consider Cooper pairs which consist of electron and neutron
(lepton and proton).The Cooper pairs existence in this case is possible due to attractive
forces considered above (attraction via polarization and via spin-spin and spin-orbital
interactions).The work in this direction is under progress (for  some estimates see \cite{I}).

Besides application to  $e^{\pm}\pi^0(K^0)$ atoms problems may be considered also
task about bound states of particles in the attractive field of the dielectric
and conducting bodies.
Due to symmetry properties the most interesting are sphere and cylinder.
In \cite{I} has been considered  also  charged particle levels in the potential
of charge particle interaction with its reflection.

For various case of conducting spheres (see \cite{LL8}) we have:
\begin{equation}
\label{A5}
V(r))=\frac{e^2R}{2(r^2-R^2)}
\end{equation}
if charge of the conducting spheres $e'$ (induced on sphere) is nonzero, and
\begin{equation}
\label{A5}
V(r))=\frac{e^2R^3}{2r^2(r^2-R^2)}
\end{equation}
if charge of conducting sphere is zero.

If radius of sphere (dielectric or conducting)
 much larger than Borh radius we obtain well known Tamm levels,the numerical results is
shown on the Fig.1-3 of the \cite{I} for case 1,2 of the ideal conducting sphere and
for case of  homogeneous dielectric sphere.

{\bf   Appendix A : Attraction from  polarization }

If we consider as first approximation neutron as homogeneous  dielectric matter inside sphere
with radius $R_n$ we can use the result of \cite{BT}. At $r>R_n$, $r<R_n$ we have \cite{BT}
respectively:
\begin{equation}
\label{A5}
V(r)=-e^2(\epsilon_{n,\bar{n}}-1)\sum^{\infty}_{l=0}\frac{l}
{l\epsilon_{n,\bar{n}}+l+1}\frac{R_n^{2l+1}}{r^{2l+2}},
\end{equation}

\begin{equation}
\label{A5}
V(r)=-e^2(\epsilon_{n,\bar{n}}-1)\sum^{\infty}_{1=0}\frac{l+1}
{l\epsilon_{n,\bar{n}}+l+1}\frac{r^{2l}}{R_n^{2l+1}},
\end{equation}

where $\epsilon_{n,\bar{n}}$ are dielectric polarizability of
neutron(antineutron).

At $r>>R_n$ only $l=1$ is essential and we obtain:
\begin{equation}
\label{A5}
V(r)=-e^2(\epsilon_{n,\bar{n}}-1)\frac{1}
{\epsilon_{n,\bar{n}}+2}\frac{R_n^3}{r^{4}},
\end{equation}
We see that at $r<R_n$ interaction is repulsive.
In  polarizability on the neutron and proton was achieved in ref.
\cite{P} (see also references therein) ($\alpha\sim 10^{-3}fm^{3}$).

Above was suggested that neutron is homogeneous dielectric sphere,
however as known neutron may be treated as nonlocal object with
with density $\rho_n(r)\sim\exp{-Mr}$.In this case we must take
into account that $\epsilon(r)$ is function on $r$.If we know
$\epsilon(r)$, in this case we can obtain electron-neutron
potential via neutron polarization by electron from Maxwell
equation:
\begin{equation}
\label{A5}
div(\epsilon(r)\vec{E})=4\pi\rho
\end{equation}
In particular because $1-\epsilon(r)\rightarrow 0$ at $r\rightarrow\infty$
by the same low as density we obtain:

\begin{equation}
\label{A5}
\epsilon(r)=(\epsilon(0)-1)\exp{(-Mr)}+1
\end{equation}
where $\epsilon(0)$ may be obtained from $e^-n$ elastic scattering
at large $\vec{q}^2$.Also $V(q)$ may be obtained from $e^-n$ elastic scattering
at all $\vec{q}^2$.In this case we can obtain the potential
$V(r)=\int V(q)\exp(i\vec{q}\vec{r})\frac{d^3q}{(2\pi)^3}$
and substitute $V(r)$ in above derived equations for electron.

From experiments on lepton-proton elastic scattering (see \cite{O} and references therein),
interaction of leptons and protons has the following form:
\begin{equation}
\label{A1}
M=eF_V(\vec{q}^2)\bar{p}\gamma_a p e\gamma_b e
\end{equation}
where  formfactor $F_V(\vec{q}^2)=\frac{M^2}{(\vec{q}^2+M^2)^2}$($M^2= 0.71 GeV^2$)
(see \cite{O} and references therein)
 are described their charge distribution in the proton.

It mean that in momentum space the potential between electron and
proton has the form $\frac{1}{q^2}\rightarrow \frac{1}{q^2}F_V(\vec{q}^2)$.

\begin{equation}
\label{A5}
V(r)=\alpha(\frac{1}{r}-\frac{\exp(-Mr)}{r}-\frac{M}{2}\exp(-Mr))
\end{equation}

Also exist potential connected with neutron formfactor:
\begin{equation}
\label{A1}
M=eF_n(\vec{q}^2)\bar{p}\gamma_a p e\gamma_b e
\end{equation}
From this formula we obtain the following short-range potential:
\begin{equation}
\label{A1}
V(r)=\int \exp{i\vec{q}\vec{r}}\frac{\alpha}{\vec{q}^2}F_n(\vec{q}^2)
\end{equation}
which decrease as $\sim \exp(-Mr)$ at large $r$.
where $F_n(0)=0$.

If  we suppose e.g. that the formfactor has the form
$F_n(t)=\frac{t^2}{(t+M^2)^2}$ (which correspond to the density
$\rho(r)=\delta(r)-M^2\frac{\exp(-Mr)}{r})$ we obtain;
\begin{equation}
\label{A5}
V(r)=-\alpha (\frac{\exp(-Mr)}{r}-\frac{M}{2}\exp(-Mr))
\end{equation}
Thus we have singular potential $\sim r^{-n}$ at $r>>r_n$.At $r<r_n<<\frac{1}{m_e}$
singular behaviour is absent.It mean the presence
of deep levels of $e^{\pm}n$ bound states. In our next paper
we will calculate it numerically.

In case of neutron the effect of neutron polarization is of order $\sim\frac{\alpha^2}{R}$
is essentially smaller than spin-spin and spin-orbital
interaction $\sim\frac{\alpha}{m_eMR^3}$.However in composite scalar-charged particle
bound states (see below) spin-spin and spin-orbital
interaction  are absent and  attraction by polarization is important.

We would like also to stress that the potential is the sum of the
potential from polarization+formfactors effect (formula ()
above)+ of spin-spin interaction and from photon-$\rho$-meson mixing .

It must be noted also exist long-range potential induced by $\pi^0\gamma\gamma$
vertex:

\begin{equation}
\label{A5}
V(q)=e^2g_{\pi n n}\frac
{1}{q^2+m^2_{\pi}}\vec{E}\vec{H}
\end{equation}
where $\vec{E},\vec{H}$ are fields of the lepton.

{\bf Appendix B: $Z^0-\gamma$ and rho-meson -photon  mixing }

We consider also potential which appear due to  $Z^0-\gamma$ mixing
via electron-positron loop. We obtain the following
result for electromagnetic field generated by $Z-\gamma$ mixing
 (only P-even part) (which act on $Q_{1,2}$):
\begin{equation}
\label{20}
eQ_{1,2}A(r)=\frac{e^2Q_{1,2}g_{2,1}(-\frac{1}{4}+\sin\theta_W^2)}{cos^2\theta_W}
\frac{1}{(2\pi)^5}\frac{4\pi^2}{3r}\int^{\infty}_1dxF(x)
(\frac{M^2\exp(-Mr)-4m^2_ex^2\exp(-2m_erx)}{M^2-4m_e^2x^2})
\end{equation}

where $g_{2,1}=(\frac{1}{2}T_{2,1}-Q_2\sin\theta_W^2)$
 P-odd part of $Z-\gamma$ mixing also give long range P-odd forces.

At $r>>\frac{1}{m_e}$ integral is suppressed by exponent $\exp(-2m_er)$
while at  $\frac{1}{M}<<r<<\frac{1}{m_e}$ we have:

\begin{equation}
\label{20}
eQ_{1,2}A(r)=\frac{4\pi^2}{3(2\pi^5)}\frac{e^2Q_{1,2}g_{2,1}(-\frac{1}{4}+\sin\theta_W^2)}{cos^2\theta_W}\frac{1}{M^2r^3})
\end{equation}

At $r<<\frac{1}{M}$ we obtain the behaviour $\sim (\frac{1}{r}+O(\alpha ln(m_er))$.

This results are also applicable for potential which appear due to  $\rho^0-meson -\gamma$
mixing via electron-positron loop.At large distances ($\frac{1}{M}<<r<<\frac{1}{m_e}$)
 this potential also fall as $\sim r^{-3}$.We can obtain the potential from
$\rho^0-meson -\gamma$ mixing taking into account that $\rho^0e^+e^-$ effective lagrangian
may be written from $\rho^0 \rightarrow e^+e^-$ decays by the following way:
\begin{equation}
\label{20}
L=\sqrt{\frac{24\pi\Gamma(\rho^0)}{M_{\rho}}}\bar{e}\hat{\rho}e
\end{equation}
(this consideration is true at $r>\frac{1}{M_{\rho}}$)
and besides interaction with nucleons is following:

\begin{equation}
\label{20}
L=g_n\bar{n}\hat{\rho}n
\end{equation}
This potential may be found from above presented formulas by the following substitution:

\begin{equation}
\label{20}
(\frac{1}{2}T_{2,1}-Q_2\sin\theta_W^2)(-\frac{1}{4}+\sin\theta_W^2)\frac{1}{cos^2\theta_W}
\rightarrow  g_n \sqrt{\frac{24\pi\Gamma(\rho^0)}{M_{\rho}}}
\end{equation}

{\bf Appendix C: Bound States of Particles with anomalous magnetic moment
(Cylidrical symmetry case)}

Below we consider cylindrically symmetric bound states
 and resonances  of particles with anomalous magnetic moments.
It is of interest to consider several special cases in particular:

1)Particle energy levels in pure electric axial field case ($\vec{E}=(E_r(r),0,0)$)
which created e.g. by
homogeneously charged cylinder.

In accordance with \cite{BT} we have:
\begin{equation}
\label{A5}
E_r=\frac{2\sigma}{r}\quad at \quad r>R
\end{equation}
and
\begin{equation}
\label{A5}
E_r=\frac{2\sigma r}{R^2}\quad at \quad r<R
\end{equation}
where $\sigma$ is density of charge of the unit of length of the cylinder.

2)Particle energy levels in pure magnetic field case ($\vec{H}=(0,H(r),0)$) which created e.g. by
homogeneously charged cylinder.

In accordance with [1] we have:
\begin{equation}
\label{A5}
H_{\phi}=\frac{2I}{r}\quad at \quad r>R
\end{equation}
and
\begin{equation}
\label{A5}
H_{\phi}=\frac{2I r}{R^2}\quad at \quad r<R
\end{equation}
where $I$ is the current in  the cylinder.

Also we consider particle energy levels in the field of solenoid ($\vec{H}=(0,0,H_z(r))$,
$H_z(r)=H\theta(R-r)$.)

Analogously may be considered situation in which
 both  electric and magnetic fields are presented.It may be for example
electric and magnetic fields created by charged particles beams.

A. Pure electric field case

The Dirac equation for particle with  anomalous magnetic moments
in pure electric axial field case takes the form \cite{LL4}:
\begin{equation}
\label{A18}
(\hat{k}-m_n+\mu(-i\vec{\alpha}\vec{E}) )\psi(k)=0,
\end{equation}
where  $\mu $-is anomalous magnetic moment,
and $\vec{E}=\vec{n}G(r)$ ($\vec{n}=(\frac{x}{r},\frac{y}{r}),r=\sqrt{x^2+y^2}$).
We find the solution in the form $\psi_{1,3}=e^{i\phi
(l-1)}e^{ip_zz}f_{1,3}(r)$,$\psi_{2,4}=e^{i\phi l}e^{ip_zz}f_{2,4}(r)$ and
in component form we obtain the following system of equations:
\begin{equation}
\label{A5}
(\epsilon-m)f_1-p_3 f_3+Q_-f_4=0
\end{equation}
\begin{equation}
\label{A5}
(\epsilon-m)f_2+Q_+f_3+p_3 f_4=0
\end{equation}
\begin{equation}
\label{A5}
p_3 f_1+ P_-f_2-(\epsilon+m) f_3=0
\end{equation}
\begin{equation}
\label{A5}
P_+f_1- p_3 f_2-(\epsilon+m)f_4=0
\end{equation}
where
\begin{equation}
\label{A5}
Q_-=i(\frac{d}{dr}+\frac{l}{r}-\mu E)
\end{equation}

\begin{equation}
\label{A5}
Q_+=i(\frac{d}{dr}-\frac{l-1}{r}-\mu E)
\end{equation}

\begin{equation}
\label{A5}
P_-=-i(\frac{d}{dr}+ \frac{l}{r}+\mu E)
\end{equation}

\begin{equation}
\label{A5}
P_+=-i(\frac{d}{dr}- \frac{l-1}{r}+\mu E)
\end{equation}
At small $r$ we find the solution in the form:
\begin{equation}
\label{A34}
f_i(r)=a_i r^{s},
\end{equation}
Substituting (51) into (43)-(50) we obtain:
\begin{equation}
\label{A35}
(\epsilon+m)(s^2-l^2-4\sigma^2\mu^2+2\mu\sigma(2s+1))a_4-p_z(-s^2-l^2-4\sigma^2\mu^2-4\mu\sigma(l+s-1))a_2=0
\end{equation}
\begin{equation}
\label{A36}
(\epsilon-m)(s^2-l^2-4\sigma^2\mu^2-4\mu\sigma s)a_2+p_z(-s^2-l^2-4\sigma^2\mu^2+4\mu\sigma s)a_2=0
\end{equation}
The determinant of this system of equations must be equal zero, and that condition
gives us the equation for defining $s$.

B. Pure magnetic field case ($\vec{H}=(0,0,H)$

The Dirac equation for particle with  anomalous magnetic moments
in pure magnetic field $\vec{H}=(0,0,H)$ case takes the form \cite{LL4}:
\begin{equation}
\label{A18}
(\hat{k}-m_n+\mu \vec{\sigma}_3 H )\psi(k)=0,
\end{equation}
In component form we obtain the following system of equations:
\begin{equation}
\label{A5}
(\epsilon-m+\mu H)f_1-p_3f_3-p_-f_4=0
\end{equation}
\begin{equation}
\label{A5}
(\epsilon-m-\mu H) f_2-p_+f_3+p_3 f_4=0
\end{equation}
\begin{equation}
\label{A5}
p_3 f_1+p_-f_2+(-\epsilon-m+\mu H) f_3=0
\end{equation}
\begin{equation}
\label{A5}
p_+f_1- p_3 f_2+(-\epsilon-m-\mu H) f_4=0
\end{equation}
where

\begin{equation}
\label{A5}
p_-=-i(\frac{d}{dr}+\frac{l}{r})
\end{equation}

\begin{equation}
\label{A5}
p_+=-i(\frac{d}{dr}-\frac{l-1}{r})
\end{equation}

In case of solenoid homogeneous magnetic
field exist only inside cylinder and we obtain:
\begin{equation}
\label{A5}
[(\epsilon^2-p^2-(m+\mu H)^2)(\epsilon^2-p^2-(m+\mu H)^2)-4\mu^2H^2(p^2-p^2_z)]\psi_4=0
\end{equation}

where $p^2=p^2_z-\frac{1}{r}\frac{d}{dr}r\frac{d}{dr}+\frac{l^2}{r}$
At $r>R$ we have:
\begin{equation}
\label{A5}
\psi_{r>R}=CK_{|l|}(\sqrt{\epsilon^2-p_z^2-m^2}r)
\end{equation}
whereas at $r<R$:
\begin{equation}
\label{A5}
\psi_{r<R}=C_{1,2}J_{|l|}((\mu^2H^2-m^2+\epsilon^2-p_z^2\pm2\sqrt{\epsilon^2-p_z^2-m^2})r)
\end{equation}
Energy levels are obtained from the condition:
\begin{equation}
\label{A5}
\frac{\psi_{r<R}'}{\psi_{r<R}}|_{r=R}=\frac{\psi_{r>R}'}{\psi_{r>R}}|_{r=R}
\end{equation}
Besides considered  above solenoid it is of interest to consider also magnetic field
created by current inside cylinder( $H=\frac{2I}{r}$ at $r>R$).Now
this consideration is in progress.

C. Electric and magnetic field case ($\vec{H}=(0,0,H_{z}(r))$)

If both  electric and magnetic fields are presented we obtain the following system of equations:
\begin{equation}
\label{A5}
(\epsilon-m+\mu H)\psi_1-p_3\psi_3+Q_-\psi_4=0
\end{equation}
\begin{equation}
\label{A5}
(\epsilon-m-\mu H)\psi_2+Q_+\psi_3+p_3\psi_4=0
\end{equation}
\begin{equation}
\label{A5}
p_3\psi_1+ P_-\psi_2-(\epsilon+m-\mu H)\psi_3=0
\end{equation}
\begin{equation}
\label{A5}
P_+ \psi_1- p_3\psi_2-(\epsilon+m+\mu H)\psi_4=0
\end{equation}

D. Electric and magnetic fields case ($\vec{H}=(0,H_{\phi},0))$)

The system of equation is following:
\begin{equation}
\label{A5}
(\epsilon-m)f_1-i\mu H f_2-p_3f_3-(p_--i\mu E)f_4=0
\end{equation}
\begin{equation}
\label{A5}
i\mu H f_1+(\epsilon-m) f_2-(p_+-i \mu E)f_3+p_3 f_4=0
\end{equation}
\begin{equation}
\label{A5}
p_3 f_1+(p_--i\mu E)f_2+(-\epsilon-m) f_3-i\mu H f_4=0
\end{equation}
\begin{equation}
\label{A5}
(p_+-i \mu E)f_1- p_3 f_2+i\mu H f_3+(-\epsilon-m) f_4=0
\end{equation}

E. Electric charge and magnetic moment joint consideration

Previously we consider neutral particles with anomalous neutral moment.If
we take into account also charge we obtain the following system of equations:
\begin{equation}
\label{A5}
A_1f_2+A_2f_4=0
\end{equation}

\begin{equation}
\label{A5}
A_3f_2+A_4f_4=0
\end{equation}
where:
\begin{equation}
\label{A5}
A_1=\frac{d}{dr}(\frac{(\epsilon-V+m)}{dk})(\frac{d}{dr}+\frac{l-1}{r})-\mu E)+
(\epsilon-V+m)( \Omega_0-\mu^2 E^2+2\mu E\frac{d}{dr}+\mu E \frac{1}{r})
\end{equation}

\begin{equation}
\label{A5}
A_2=p_z( \Omega_0+\mu^2 E^2+2\mu E\frac{d}{dr}+\mu E'+\mu E\frac{2l-1}{r}+
\frac{dk^{-1}}{dr}(\frac{d}{dr}+\frac{l-1}{r}+\mu E)
\end{equation}

\begin{equation}
\label{A5}
A_3=(-\epsilon+V+m)( \Omega_0-\mu^2 E^2-2\mu E\frac{d}{dr}-\mu E'-\mu E \frac{1}{r})-
k\frac{d}{dr}(\frac{(\epsilon-V+m)}{dk})(\frac{d}{dr}+\frac{l-1}{r})+\mu E)
\end{equation}

\begin{equation}
\label{A5}
A_4=p_z( -\Omega_0-\mu^2 E^2+2\mu E\frac{d}{dr}+ \mu E'+ \mu E\frac{1}{r}+
\frac{1}{2}\frac{dk^{-2}}{dr}(\frac{d}{dr}+\frac{l-1}{r}- \mu E)
\end{equation}

If charge is zero (i.e. at $V=0$) we obtain previous formulas (43)-(50).

The author express his sincere gratitude to H.Asatryan, E.Gazazyan, G.Grigoryan, S.Elbakyan,
S.Harutyunyan, R.Pogossyan, E.Prokhorenko for helpful discussions.


\begin{thebibliography}{99}
\bibitem{LL4}L.D.Landau,E.M.Lifshits V.4 "Quantum Electrodynamics"
\bibitem{O}L.B.Okun "Quarks and Leptons"
\bibitem{P}Phys.Rev.D50(1994)Part 1 Review of particle
properties (see also references therein)
\bibitem{BT}V.V.Batigin,I.N.Toptigin "Collection of tasks on electrodynamics"
(Moscow,1970)
\bibitem{I}R.A.Alanakyan YERPHI-1538(12)-99
\bibitem{LL8}L.D.Landau,E.M.Lifshits V.8 "Electrodynamika sploshnikh sred"
\end{thebibliography}
\end{document}